\documentclass[runningheads]{svmult}

\usepackage{makeidx}   
\usepackage{graphicx}  
\usepackage{subeqnar}  
\usepackage{multicol}  
\usepackage{physprbb}  
\makeindex             

\begin{document}
\title*{Massive Variability Search and Monitoring by OGLE and ASAS}
\toctitle{Focusing of a Parallel Beam to Form a Point
\protect\newline in the Particle Deflection Plane}
%
%
\titlerunning{OGLE and ASAS}
%
\author{Bohdan Paczy\'nski}
\institute{Princeton University, Princeton NJ 08544, USA}

\maketitle              

\begin{abstract}
OGLE and ASAS are long term observing projects operated by the Warsaw
University Observatory at the Las Campanas site in Chile.  OGLE is currently
monitoring almost 200 million stars in the Galactic Bulge and the Magellanic
Clouds, and has detected so far almost 1,000 events of gravitational
microlensing with the dedicated 1.3-meter telescope.  ASAS uses several very
small instruments to monitor all southern sky for variability down to
approximately 14 magnitude.  A total of almost 300 thousand variable stars
were discovered so far by the two projects, and all photometric data is
available on the WWW.  Both projects aim at real time recognition and 
verification of all new phenomena in the sky.  OGLE is likely to discover
planets by 2003 and stellar mass black holes by 2004-2005.  All OGLE and ASAS
data is made public domain as soon as possible, and may be used by a
Virtual Observatory.
\end{abstract} 

\section{OGLE - Past and Present} 

OGLE (Optical Gravitational Lensing Experiment, the leader: Andrzej Udalski)
is a long term project, with a gradual expansion of its capability.
The dedicated 1.3 meter telescope is located at the
Las Campanas Observatory in Chile, a site owned and operated by the
Carnegie Institution of Washington.  OGLE is operated by the Warsaw
University Observatory.  Information about it may be found at:
\vskip 0.2cm
\centerline{OGLE: \hskip 1.0cm http://sirius.astrouw.edu.pl/\~~ogle/}
\centerline{OGLE: \hskip 1.0cm http://bulge.princeton.edu/\~~ogle/ \hskip 0.05cm}
\vskip 0.2cm

OGLE-III is the current, third stage of the project.  The observations
are done with a mosaic $ 8K \times 8K $ CCD camera built by 
A. Udalski.  OGLE-III begun in May 2001.  In the 2002 Galactic Bulge
season almost 400 candidate microlensing events were discovered:

\centerline{OGLE-EWS: \hskip 1.0cm http://sirius.astrouw.edu.pl/\~~ogle/ogle3/ews/ews.html}
\noindent
A search for planetary transits was conducted on 32 nights during a 45 day
interval in 2001.  A total of 59 stars were identified, for which low depth
flat bottom transits with orbital periods shorter than 10 days were found 
(Udalski et al. 2002a,b).  Some of these are likely caused by 
`hot Jupiters', others by brown and red dwarfs.

The analysis of OGLE-II stage (Udalski, Kubiak \& Szyma\'nski 1997), 
which covered years 1997-2000, is advanced, and almost all data is in public
domain.  This includes catalogs of I, V, B magnitudes and positions for
a total of almost 40 million stars in the Galactic Bulge, and in the
Magellanic Clouds (Udalski et al. 2002c, 2000b, 1998a), catalogs of over
500 microlensing events (Udalski et al. 2000a, Wo\'zniak et al. 2001),
catalogs of over 270 thousand variable stars (Zebru\'n et al. 2001,
Wo\'zniak et al. 2002), and in particular thousands of eclipsing binaries in
the SMC (Udalski et al. 1998a) and Cepheids in the LMC and SMC (Udalski et al.
1999a,b).  In addition OGLE astrometry provided proper motion measurements of
thousands of stars (Soszy\'nski et al. 2002, Sumi et al. 2002), leading to the
discovery of streaming motion (rotation) in the Galactic Bar.

Of particular interest are special microlensing events.  OGLE-2000-BUL-43
was found to have a spectacular parallax effect (Soszy\'nski et al. 2001).
Even more dramatic was OGLE-1999-BUL-19, the first event with multiple
peaks in its apparent brightness caused by the Earth's orbital motion
and a very long time scale: $ t_E = R_E / V = 372 $ days (Smith et al. 2002).  
Even longer time scales were found for OGLE-1999-BUL-32 (Mao et al. 2002,
$ t_E = 641 $ days and OGLE SC\_5 2859 (Smith 2002, $ t_E = 551 $ days).
The last two events are likely caused by massive lenses, probably stellar 
mass black holes.

\section{ASAS - Past and Present}

ASAS (All Sky Automated Survey, the leader: Grzegorz Pojma\'nski) is a long
term project, with gradual expansion of its capabilities.  Currently it
has 4 small instruments, with apertures of 2 cm, 7 cm, 7 cm, and 20 cm, all
located at the
Las Campanas Observatory in Chile, at a site owned and operated by the
Carnegie Institution of Washington.  ASAS is operated by the Warsaw
University Observatory.  Information about it may be found at:
\vskip 0.2cm
\centerline{ASAS: \hskip 1.0cm http://www.astrouw.edu.pl/\~~gp/asas/asas.html}
\centerline{ASAS: \hskip 1.0cm http://archive.princeton.edu/\~~asas/ \hskip 1.6cm}
\vskip 0.2cm

All four instruments use as detectors Apogee $ 2K \times 2K $ CCD cameras.
The two instruments with 7 cm aperture cover all sky every two nights in
standard V and I filters down to about 14 magnitude.  A total of over 6 
thousand variables was discovered so far by ASAS 
(Pojma\'nski 1998, 2000, 2002), and all photometric data are on the WWW.

All instruments are fully robotic, with some support provided by the OGLE
observers.

\section{Future of OGLE and ASAS}

Both projects aim at real time data processing, and in particular at real
time recognition and verification of any new events in the sky, among them
microlensing events, supernovae, novae, dwarf novae, stellar flares, GRB
afterglows, etc.

Two specific goals of OGLE are: a firm detection of microlensing by planets,
and by stellar mass black holes.  So far stellar mass black holes were 
suggested as lensing masses for several long microlensing events
(Bennett et al. 2002, Mao et al. 2002, Smith 2002), but evidence is not
conclusive.  No definite planetary microlensing event has been discovered
so far (cf. Bennett et al. 1999, Albrow et al. 2000, Gaudi et al. 2002).

Inspecting almost 400 microlensing events reported by OGLE-III EWS (Early
Warning System) in 2002 Jaroszy\'nski \& Paczy\'nski (2002) noticed that
the event OGLE-2002-BLG-055 has a single data point deviating from otherwise
smooth microlensing light curve by 0.6 magnitudes, while the photometric
errors were about 0.01 magnitude.  Dr. Udalski kindly examined the CCD
image and found that there was nothing wrong with it, i.e. the bright
point appears to be
real.  As nearby data points show no obvious departure from a smooth
light curve a plausible interpretation of the phenomenon is in terms of
a binary lens with a very extreme mass ratio.  While a unique value
of the mass ratio cannot be determined from so sparse time coverage,
a good fit to all data was obtained for the mass ratio of 0.01 and 0.001,
indicating a planetary mass companion to a stellar mass lens.  Obviously,
it is not possible to make a strong claim based on a single data point,
but a modest modification of the future OGLE observing procedure will
provide a far better coverage of future planetary events.  There are
typically only several hundred stellar microlensing events unfolding at 
any given time, compared
to over 150 million stars monitored on a given night.  Small CCD
sub-frames covering known events can be processed within minutes of data
acquisition.  When an anomalous data point is noticed the observation
of the field will be repeated.  If the anomaly is confirmed the field
will be observed every 30 or 60 minutes, to provide a good coverage of
the rapidly changing brightness, and allowing a unique determination
of the mass ratio.  We expect that OGLE-III EWS system will have this
capability by the spring of 2003 (A. Udalski, private communication), and
it is likely that the first definite planetary microlensing event will
be detected in 2003.

At the opposite end of the lens mass spectrum are the very long duration
events, which are plausible candidates for stellar mass black holes.
Currently two independent lens parameters can be determined for long 
events with the OGLE-III photometry:
the event time scale $ t_E $ and the magnitude of the parallax effect.
One additional parameter: the angular separation between the two images,
or the astrometric shift in the combined light centroid is needed to
determine uniquely the lens mass.  It will be possible to make
such measurements with the future VLT Interferometer
(Delplancke et al. 2001, Segransan et al. 2002), or with the existing HST.
A massive lens like the one associated with OGLE-1999-BUL-32 had the
two images separated by several milli arc seconds (Mao et al. 2002, eq. 9),
leading to the centroid motion of a comparable amount.  This astrometric
effect is within easy reach of the HST (e.g. Benedict et al. 2002).
The long events last several years, and with the OGLE-III data rate there
will be several of them unfolding at any given time within several degrees 
of the Galactic center.  Very likely the VLTI and/or  HST observations
may be scheduled in advanced, with no need for the TOO (Target of 
Opportunity) mode of operation.  Little is known about the number and the
distribution of stellar mass black holes in the Galaxy, and the range of
their masses.  OGLE is likely to lead to the first definite mass determination
in 2004 - 2005.

We expect that OGLE as well as ASAS will expand their capability in the future.
There are no definite and specific plans for the expansion, but there is
a general idea how to proceed.  A natural step to OGLE-IV would be another
1.3 - 1.8 meter telescope with a CCD camera with a total of $ \sim 1 $ giga
pixels, i.e. an instrument with the data rate 16 times higher than current 
OGLE-III.  The data rate is the single most important parameter for the search
and/or monitoring projects like OGLE and ASAS.  For ASAS we expect more 
frequent all sky coverage, with a more gradual increase in its depth.  A 
virtue of bright transients is that they may be followed up for a longer 
time, in greater detail and with more instruments than faint transients can.
Obviously, bright transients are rare, hence the need for all sky coverage 
and a frequent time sampling.

It is a pleasure to acknowledge the support by NSF grants AST 9820314
and AST 0204908, and NASA grant NAG5-12212.

To appear in: ``Towards an International Virtual Observatory'', June 2002,
Garching bei M\"unchen (Germany), eds. G\'orski K. M. et al, ESO conference
series.


\end{document}